\documentclass{aa}
\usepackage{graphicx}
\begin{document}
   \title{Non-circular rotating beams and CMB experiments}

   \author{J.V. Arnau
          \inst{1}
          A. M. Aliaga
          \inst{2}
          \and
          D. S\'aez\inst{2}}

   \offprints{D. S\'aez}

   \institute{Departamento de Matem\'atica Aplicada,
              Universidad de Valencia,
              46100 Burjassot, Valencia, Spain\\
         \and
             Departamento de Astronom\'{\i}a y Astrof\'{\i}sica, 
             Universidad de Valencia, 46100 Burjassot, Valencia, Spain\\
             }

   \date{Received August 2001; accepted November 2001}

   \abstract{This paper is concerned with small angular scale experiments 
for the observation of cosmic microwave background 
anisotropies. 
In the absence of beam, the effects of partial coverage and pixelisation 
are disentangled and analyzed (using simulations). Then,
appropriate maps involving
the CMB signal plus 
the synchrotron and dust emissions from the Milky Way are
simulated, and
an asymmetric beam --which turns following different strategies--
is used to smooth the simulated maps.
An associated circular beam is defined to estimate the 
deviations in the angular power
spectrum produced by beam
asymmetry without rotation and,
afterwards, the deviations due to beam 
rotation are calculated. For a certain large coverage,
the deviations due to pure asymmetry and asymmetry plus rotation
appear to be very systematic 
(very similar in each simulation). 
Possible applications of the main results of this paper 
to data analysis in large coverage experiments --as PLANCK--
are outlined.
   \keywords{cosmology: cosmic microwave background, 
cosmology:theory---
large-scale structure of the universe---methods:numerical
---methods:data analysis}
   }

   \maketitle
%

\section{Introduction}

Many experiments are being designed for the observation
of the Cosmic Microwave Background (CMB) anisotropies.
From the maps of a given experiment operating with 
a non-circular (asymmetric) 
rotating beam, a certain angular power spectrum ($C_{\ell}$ quantities)
can be extracted. Different rotations 
can lead to distinct $C_{\ell}$ coefficients and, 
the question is: How different are these coefficients? In other
words, how relevant is the effect of the rotation strategy 
on the resulting angular power spectrum? 

In a previous paper (Arnau \& S\'aez 2000), 
it was shown that, in the absence of
rotation and when the level of instrumental noise is low enough, 
the effect of a non-circular beam can be subtracted
--namely, the beam can be deconvolved-- using the Fourier transform.
This subtraction can be performed in such a way that the resulting 
spectrum, after deconvolution, 
is very similar to the true one. That is possible if the
number of pixels inside the beam, $N_{\mbox{in}}$, 
is not too great. Indeed,
$N_{\mbox{in}}$ cannot be much greater than 10; 
however, if the beam rotates, 
the deconvolution
is not possible. Nobody has described either the importance of
beam rotation or a method to eliminate its effects.
The main goal of this paper is the estimation of the effects due to
rotation.
In Arnau \& S\'aez 2000 (and also in S\'aez, Holtmann \& Smoot, 1996, 
and S\'aez\& Arnau, 1997). 
a sort of modified angular power spectrum 
was used. Here, we extract the standard $C_{\ell}$ quantities
from a certain number of squared patches of the sky.
Recently, Wu et al. (2001) have proposed a method for data 
analysis in the case of asymmetric beams. This method
is based on an optimal 
circular beam associated to the asymmetric one.
The effects of beam rotation are not studied at all by these
authors. 

Although our methods apply to CMB anisotropy experiments in
general, we will pay particular attention to PLANCK mission
(scheduled by ESA for 2007). As it was emphasized in Burigana et al.
(1998), {\em beam responses are typically nonsymmetric for 
detectors de-centred from the telescope focus}. Taking into account 
that CMB anisotropy experiments require observations at 
different frequencies, various detectors are necessary, which must
be distributed as close as possible from the focus; for instance, 
in the PLANCK 
mission, around one hundred of detector (bolometers and radiometers)
must be distributed in the focal plane. If the focal plane rotates
(rotation of the telescope around the spin axis),
the beams do. The effect of this rotation 
deserves attention. 
Furthermore, there are various identical
detectors for each frequency, which are located at different
positions in the focal plane and, consequently, the deformations 
of these beams would be different (identical) if they are 
located at different (the same) distances from the optical focus;
nevertheless, even for identical deformations, the orientations 
of the resulting asymmetric beams would be different.
The motion of the line of sight through the sky also produces a 
beam asymmetry. The effective beam diameter
$\theta_{_{\mbox{FWHM}}}$ appears to be enhanced in the direction 
of this motion (see Hanany, Jaffe \& Scannapieco 1998). This
small effect is due to the beam displacement during the
measurement process. It is not taken into account  
in this paper.

Beam rotation depends on the particular experiment
under consideration. Given a pixelisation, the beam centre 
points towards a given pixel a certain number 
of times, $N_{\mbox{p}}$, and, then,
the temperature assigned to this pixel is an average of the 
temperatures corresponding to each of the $N_{\mbox{p}}$ measurements. 
The fact that measurements from various beam orientations 
are averaged could be important.

In the case of PLANCK mission,
a rough estimate of number $N_{\mbox{p}}$ is given in S\'aez \& Arnau (2000).
Here, it is worthwhile to improve a little on that estimate.
The satellite has been designed in such a way that: (i) it
will cover the full sky in seven months, with a coverage which
can be considered as uniform in most part of the sky, (ii) 
its line of sight will move around a big circle on the sky completing 
a turn each minute and, (iii) it will move around the same circle for two 
hours (120 turns). On account of these facts, 
if the pixel 
size is $\Delta$ and 
the angle subtended by the motion of the line of sight 
between two successive measurements is $\Delta \alpha =
\zeta \theta_{_{\mbox{FWHM}}}$, where $\theta_{_{\mbox{FWHM}}}$ is
the beam diameter, then, the average number of measurements 
per pixel (in a seven months observing period) is
$N_{\mbox{p}} = 42 \Delta^{2} / \zeta \theta_{_{\mbox{FWHM}}}$,
where all the angles are given in arc-minutes 
(see S\'aez \& Arnau, 2000, for comparison) and, furthermore,
the average number of measurements per pixel performed while 
the line of sight 
turns 120 times around a given circle is 
$N_{\mbox{pc}} = 120 \Delta / \zeta \theta_{_{\mbox{FWHM}}}$.
From these formulae, it follows that the average number 
of circles passing by a pixel --during seven months--
is $N_{\mbox{c}} = N_{\mbox{p}}/N_{\mbox{pc}} \simeq \Delta/3$, this result is 
consistent
with the fact that, for a given observational strategy, 
the number $N_{\mbox{c}}$ is expected to be dependent only 
on the pixelisation. Of course, it is independent on 
beam asymmetry.
The number of measurements corresponding to different orientations 
could be important for the effect we are looking for, 
which is produced by the rotation of asymmetric beams. 
The larger the
pixel size, the better the situation (the greater $N_{\mbox{c}}$). 

Since the detectors are rigidly 
attached to the focal plane, any beam has almost 
the same orientation each time it crosses a given pixel 
during its motion (120 turns) along a given circle;
however, this orientation changes from circle to circle.
From the above comments and estimates, it follows that 
the average number of measurements per pixel 
corresponding to different beam orientations is
$N_{\mbox{c}}$. If the full sky is covered two times and, 
the second coverage is not identical to the first one, 
this average number would be 
$2 \Delta/3$. For $5^{\prime } < \Delta < 10^{\prime}$,
this number ranges from 3.3 to 6.6. 
Nevertheless, there are various detectors in the focal
plane for each frequency and, by assuming that all the beams
have the same shape but different orientations, 
the above $N_{\mbox{c}}$ number can be multiplied by the number of
beams.

\section{Beam}

Our asymmetric beam is assumed to be of the form:
\begin{equation}
W(\theta, \phi) = W_{_{\mbox{N}}} 
e^ {\left[ - \frac {(\theta - \theta^{\prime})^2}
{2 \sigma_{\theta}^{2}} - \frac {(\phi - \phi^{\prime})^2}
{2 \sigma_{\phi}^{2}} \right]} 
\label{asb}
\end{equation}
as in Burigana et al. (1998). It is hereafter called an elliptical
beam.
Estimations of the effects due to 
asymmetry require the definition of an associated
circularly symmetric beam for comparisons.  Here the  
associated beam is defined as follows:
given a circle of radius $R=(\theta^{2} + \phi^{2})^{1/2}$, the
weight $W(R)$ corresponding to the circular beam is the average 
value of the weights $W(\theta, \phi)$ assigned 
by the elliptical beam to the points of the circle; namely,
\begin{equation}
W(R)=\langle W(\theta, \phi) \rangle_{_{\mbox{C}}} \ .
\label{sb}
\end{equation}
There are arguments suggesting that the
circular beam given by Eq. (\ref{sb}) is 
very appropriate. Let us discuss this point
in some detail.
Suppose an uniform temperature field $T(\theta , \phi)=constant$.
Place the centre of the elliptical beam at point P with 
arbitrary orientation and, then,  
consider various circles centred at P 
with radius $R_{i}$. The averages inside each of these circles 
performed with the asymmetrical beam and with the associated 
circular one are
identical (except for errors produced by discretisation in
Eqs. (\ref{asb}) and (\ref{sb})). 
These averages coincide whatever the radius $R_{i}$ 
and the orientation of the 
asymmetric beam may be. 
With any other circular beam, 
there would be deviations between the measurements 
of both beams, which would depend on the spatial scale (on $R_{i}$), 
that feature does not seem to be appropriate
because scale dependent differences 
would appear as an artifact.
It could be argued that we are considering a very special 
uniform field which is very different from the true 
CMB maps; nevertheless, 
in spite of the fact that the cosmological signal 
has fluctuations at different scales,
this signal is expected to be an homogeneous and 
isotropic statistical field and, consequently, 
the above associated beams seem to be 
appropriate in order to avoid artifacts
after measurements and {\em averages}; namely, from a
statistical point of view. 
The differences between the angular power spectra 
after averaging with the two associated beams are not artifacts due to 
a bad association, but small differences produced by other 
numerical or physical reasons. For example, as a result of 
pixelisation 
(a type of discretisation) beams (\ref{asb}) and (\ref{sb})
deviate.
It is worthwhile to notice that
the associated circular beam can be easily
obtained --using Eq. (\ref{sb})--
whatever the asymmetric beam may be and, also, that this 
association does not depend on the features of the maps to be 
smoothed by the beam (which are not known {\em a priori}).
In the case of the
beam defined by Eq. (\ref{asb}), the associated
circular beam appears to have the following form:
\begin{eqnarray}
W(R)=  \nonumber \\
=\frac {2W_{_{\mbox{N}}}}{\pi} e^{-R^{2} \sigma_{\phi}^{-2}/2} 
\int_{0}^{\pi /2}
e^{[R^{2} (\sigma_{\theta}^{-2}-\sigma_{\phi}^{-2})\sin^{2} \xi]/2}
d \xi  \ ,   
\label{wr}
\end{eqnarray}
which is used below in numerical estimates. 

The total signal --measured in a certain frequency-- also involves
components which are not statistically homogeneous and isotropic;
for instance, dust and synchrotron radiations from our galaxy; 
nevertheless, we cannot use two different circular associated beams,
but only one, and the fact that we are particularly interested in the
CMB signal strongly suggests the use of beam (\ref{sb}).


\section{Map making algorithm and power spectrum estimator}

We are concerned with a $\Lambda CDM$ model, which is 
a standard inflationary (flat) one with cold dark matter,
having 
$\Omega_{\mbox{b}}=0.05$, $\Omega_{\mbox{d}}=0.25$, $\Omega_{\Lambda}=0.7$,
and $h=0.65$,
where quantities 
$\Omega_{\mbox{b}}$, $\Omega_{\mbox{d}}$ and  $\Omega_{\Lambda}$ stand for the
density parameters corresponding to baryons, dark matter, and 
vacuum energy densities, respectively, and
quantity  h is the reduced Hubble constant.
In this model, 
the CMB temperature 
is a Gaussian homogeneous and isotropic statistical 
two dimensional field. In such a case,  
a certain method proposed by  
Bond \& Efstathiou (1987) can be used to make 
the $18^{\circ} \times 18^{\circ}$ maps 
used in this paper. This method is based on the
following formula: 
\begin{equation}
\frac {\delta T}{T} = \sum^{N}_{s_{1},s_{2} = -N} D(\ell_{1},
\ell_{2})e^{-i(\theta \ell_{1} + \phi \ell_{2})} \ ,
\label{mm}
\end{equation}
where $\ell_{1} = 2 \pi s_{1} / \Lambda$, 
$\ell_{2} = 2 \pi s_{2} / \Lambda$, and $\Lambda $ stands for  
the angular size of the square to be mapped.
This equation defines a Fourier transform from 
the position space ($\theta$, $\phi$) to the momentum space
($\ell_{1}$, $\ell_{2}$).    
The Gaussian quantities $D(\ell_{1},\ell_{2})$ have zero mean, 
and 
their variance is proportional to $C_{\ell}$, 
where $\ell = (\ell_{1}^{2} +
\ell_{2}^{2})^{1/2}$. Since $\delta T / T$ is real,   
the relation $D(- \ell_{1},- \ell_{2}) =D^{\ast}(\ell_{1},\ell_{2})$
must be satisfied. From given $C_{\ell}$ coefficients,  
the above  
$D(\ell_{1},\ell_{2})$ quantities  can be easily calculated and, then,
according to Eq. (\ref{mm}),
a Fourier transform leads to the map.
S\'aez, Holtmann \& Smoot (1996) used this map making 
algorithm to get very good simulations of 
$20^{\circ} \times 20^{\circ}$ squared regions.   

In the case of small squared maps,
the above map making
method suggests a power spectrum estimator. 
Given one of these maps $\delta T / T (\theta , \phi) $, 
an inverse Fourier transform leads to quantities 
$D(\ell_{1},\ell_{2})$ and, then, the average 
$\langle |D(\ell_{1},\ell_{2})|^{2} \rangle$
can be calculated on the circumference 
$\ell^{2} = \ell_{1}^{2} + \ell_{2}^{2}$. 
Some interpolations are necessary to get 
the $D(\ell_{1},\ell_{2})$ values at the points located on  
the circumference. 
The resulting average is proportional to $C_{\ell}$, where $\ell$
is the radius of the circumference.

Another map making algorithm and a different 
power spectrum estimator have been also used 
for comparisons. A few comments about these methods, which play
an auxiliary role in this paper, are worthwhile.  

The effect of partial coverage --without 
considering pixelisation-- was studied by Scott,
Srednicki \& White (1994). In an experiment 
covering a fraction, $f_{\mbox{sky}}$, of
the sky, 
these authors showed that the sample variance is just 
the cosmic one enhanced 
by the factor $f_{\mbox{sky}}^{-1}$.
The meaning of the cosmic variance was discussed 
in detail by L. Nox (1995). 
From these papers it follows that, in an experiment
with partial coverage,  
the deviations of the estimated angular power spectrum 
$C_{\ell}^{\mbox{est}}$ with respect to the average $C_{\ell} = \langle
C_{\ell}^{\mbox{est}} \rangle$ obey the relation
$\langle (C_{\ell}^{\mbox{est}}-C_{\ell}) (C_{\ell^{\prime}}^{\mbox{est}}
-C_{\ell^{\prime}})\rangle  = [2/(2 \ell +1)f_{\mbox{sky}}]C_{\ell}^{2} 
\delta_{\ell \ell^{\prime}}$ and, consequently,
the relative errors of the resulting $C_{\ell}$
quantities are:
\begin{equation}
\frac {\Delta C_{\ell}} {C_{\ell}}
 = \left[ \frac {2} {(2 \ell + 1)f_{\mbox{sky}}} \right]^{1/2} \ .
\label{knox}
\end{equation}

%
   \begin{figure}[!h]
   \centering
   \includegraphics[width=8.5cm]{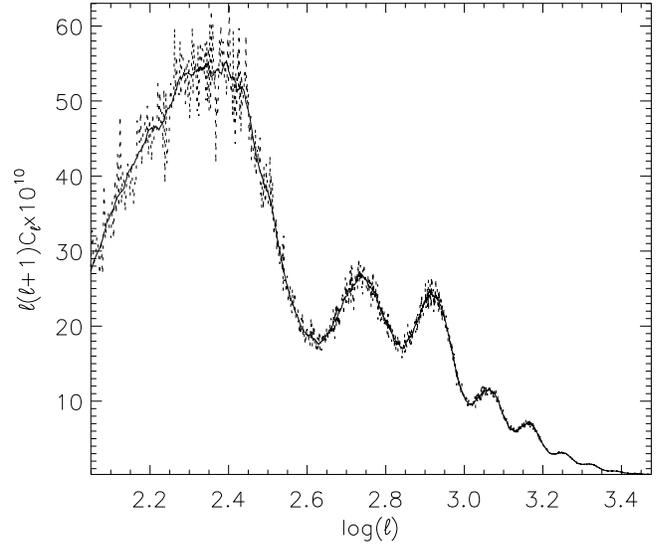}
      \caption{Dotted line shows the angular power spectrum 
   $C_{\ell}^{\mbox{est}}$, which involves high frequency oscillations
   (sample variance), whereas solid line  
   approaches the true
   spectrum $C_{\ell} = \langle C_{\ell}^{\mbox{est}} \rangle$}
         \label{nFig1}
   \end{figure}

In Fig.~\ref{nFig1}, two angular power spectra are displayed.
One of them (dotted line) has been directly 
calculated using the HEALPIX (Hierarchical Equal Area isoLatitude
Pizelisation) package. First, a full sky simulation has been made 
using the code SYNFAST
with the following inputs: (1)
the $C_{\ell}$ coefficients of the $\Lambda CDM$ model under
consideration, and (2) a pixel size 
$\Delta = 3.435$ ($N_{\mbox{side}}$=1024) and,
then, the resulting map has been analyzed using the code ANAFAST, which 
gives the angular power spectrum. We have excluded an equatorial 
band in the analysis and, consequently, the resulting spectrum  
corresponds to two polar regions
covering a part of the sky with $f_{\mbox{sky}} \simeq 0.39$.
Since ANAFAST gives the quantities $C_{\ell}^{\mbox{est}}$ defined 
above, the dotted line shows high 
frequency oscillations (sample variance). 
We can now use some appropriate numerical method 
to estimate the quantities
$C_{\ell} = \langle C_{\ell}^{\mbox{est}} \rangle$. A simple 
method suffices for us, we have taken 
$\langle C_{\ell}^{\mbox{est}} \rangle = \frac {1}{21} 
\sum_{i= \ell - 10}^{\ell + 10} C_{i}^{\mbox{est}}$. The resulting smooth 
spectrum is that of the solid line. The high frequency 
oscillations due to partial coverage have been
partially subtracted from the spectrum displayed in 
the dotted line; thus, the solid line shows a spectrum which only 
involves a residual part of the
effect due to partial coverage. This residual effect 
should be also proportional to $f_{\mbox{sky}}^{-1}$.  

The power spectrum estimator based on the Fourier transform
directly approaches quantities
$C_{\ell} = \langle C_{\ell}^{\mbox{est}} \rangle$; namely,
the spectrum given by this estimator does not involve 
high frequency oscillations, but only  
a residual effect of partial coverage
proportional to $f_{\mbox{sky}}^{-1}$ (see below). 
This spectrum is comparable to 
that of the solid line of  
Fig.~\ref{nFig1}. Hereafter, only this type of spectra are
considered and, consequently, we are concerned with
the Residual Effect of Partial Coverage (REPC). 
Using these spectra, 
results obtained from squared patches can be compared
with results corresponding to the HEALPIX package.  

\section{Results}

We first consider simulated maps containing only 
the CMB signal.  
The CMB angular power spectrum is used to built up
$18^{\circ} \times 18^{\circ}$ maps which have either 256
or 128 nodes per edge. 
For 256 (128) nodes, the pixel size is
$\Delta = 4.7^{\prime}$ ($\Delta = 9.4^{\prime}$) and, in the 
PLANCK case, the
average number of measurements per pixel corresponding to 
distinct orientations of a given
beam is $N_{\mbox{c}}=3.13$ ($N_{\mbox{c}}=6.26$).
Hereafter, the elliptical beam is --implicitly-- assumed to be an 
asymmetric one 
of the form (\ref{asb}) with
$\sigma_{\theta}=6^{\prime}$ and
$\sigma_{\phi}=10^{\prime}$, 
excepting a few cases where other beams 
are explicitly defined. 
Moreover, any set of $n$ $18^{\circ} \times 18^{\circ}$
simulated maps is called a $n$-simulation.

\subsection{Pixelisation and partial coverage}

\begin{figure*}
   \centering
   \includegraphics{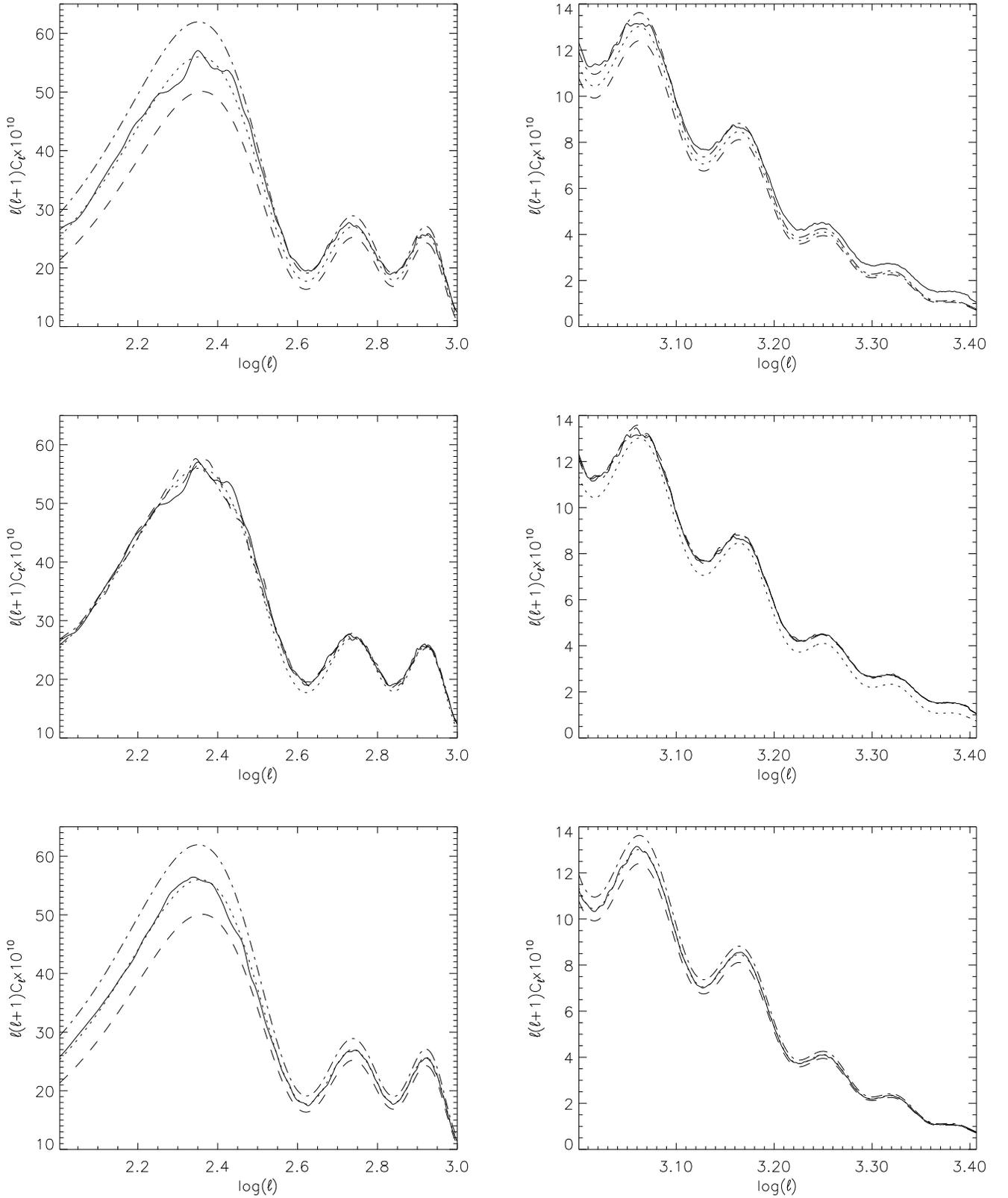}
   \caption{Left (right) top panel shows four spectra
   for $\ell < 1000$  ($\ell > 1000$): Dotted lines 
   display the theoretical spectrum used to simulate
   $18^{\circ} \times 18^{\circ}$ maps with 256 nodes
   per edge and no beam. Solid line is the spectrum obtained from 
   50 simulated maps (50-simulation). 
   Dashed (dotted-dashed) lines corresponds
   to $C_{\ell} - \Delta C_{\ell}$ ($C_{\ell} + \Delta C_{\ell}$), 
   where $\Delta C_{\ell}$ is given by Eq. (\ref{knox}).
   Middle panels show three spectra: Dotted line is 
   the theoretical spectrum and solid and dashed lines
   are the spectra obtained from two independent 50-simulations.
   In bottom panels, dotted, dashed and dotted-dashed lines
   show the same spectra as in top panel, but solid line 
   is the spectrum obtained from simulations after subtracting 
   (see text) the 
   systematic effect due to pixelisation}
              \label{Fig1}%
    \end{figure*}

First of all, the superposition  
of the REPC and the pixelisation effect 
is estimated in the absence of any beam.
The REPC depends on the coverage 
and the pixelisation effect depends 
on the pixel size $\Delta$.
Pixelisation is a mathematical
discretisation and, consequently, any 
discretised mathematical formula used in our numerical
procedures may induce an effect, which
could appear 
as a deviation of the resulting angular power spectrum 
with respect to the true one.
Discretised 
mathematical formulae can be involved in the
method to extract the spectrum from 
the maps and even in the simulation procedure.
That makes no possible
the definition of a pixelisation effect depending
only on the pixelisation itself; in 
each case (signal plus mathematical methods of 
simulation and analysis), the effect of discretisation
(pixelisation) must be estimated. We do that below 
(for squared patches) and
the resulting effect appears to be very
systematic, namely, it appears to be almost the 
same in any simulation. The same occurs when the 
HEALPIX package is used.
Of course, the spectra obtained from our simulations include both
the REPC and the pixelisation effect. 
In Fig.~\ref{Fig1}, all the spectra are obtained 
from fifty $18^{\circ} \times 18^{\circ}$
maps (50-simulation) 
having 256 nodes per edge, namely from a partial sky coverage with
$f_{\mbox{sky}} \simeq 0.39$ and $\Delta = 4.7^{\prime}$.
The effect of pixelisation becomes dramatic for
$\ell > \pi / \Delta$, where $\Delta$ is the pixel size
and, consequently, the spectra are only showed
for $\ell < \ell_{\mbox{max}} = \pi / \Delta=2550$.
Each spectrum is displayed in two panels
(left and right) to make visible some details.
In the top panel of Fig.~\ref{Fig1}, 
the solid line corresponds to the spectra obtained from 
a 50-simulation, 
the dotted line 
shows the true $C_{\ell}$ coefficients
used in the simulations and,   
finally, the dashed and dotted-dashed lines 
correspond to $C_{\ell} - \Delta C_{\ell}$ and 
$C_{\ell} + \Delta C_{\ell}$, where $\Delta C_{\ell}$
is given by Eq. (\ref{knox}).
The effect produced by pixelisation 
is expected to be dominant for large $\ell$-values
and this effect appears always mixed with the REPC.
For $\ell > 1600$ (see top right panel), the
solid line is well
outside the region limited by $C_{\ell} - \Delta C_{\ell}$ 
and $C_{\ell} + \Delta C_{\ell}$, which suggests
that the effect due to pixelisation 
is clearly dominant. For $1600 > \ell > 100$, there is
a mixing of the two effects under consideration
(which are disentangled below). Finally, the largest values
of the REPC should be found for
$100 < \ell < 300$ (see top and middle left panels).
   
In the middle panel of Fig.~\ref{Fig1},
the dotted line 
shows the true $C_{\ell}$ coefficients, whereas
each of the other two lines corresponds to the spectrum 
obtained from a different 50-simulation (these two spectra 
are denoted $C_{\ell}^{1}$ and $C_{\ell}^{2}$ in this
Section). It is
noticeable that the deviations of these two lines 
with respect to the dotted line (true spectrum)
are very 
similar. This fact indicates that --for our signal
and numerical procedures--, the addition of the REPC and
the pixelisation effect has
a dominant systematic part. 
The average deviations are the quantities
$D_{\ell} = C_{\ell} - (C_{\ell}^{1} 
+ C_{\ell}^{2})/2$, and the quantities
$\tilde{C}_{\ell} = C^{1}_{\ell} - D_{\ell}$ 
(hereafter named the corrected spectrum) 
are given in the bottom panels of Fig.~\ref{Fig1}
(solid line), where we
can verify that quantities
$\tilde{C}_{\ell}$ are very similar to the
true $C_{\ell}$ coefficients (dotted line) all along the interval 
(100, $l_{\mbox{max}}$). The dashed and doted-dashed
lines of this panel have the same meaning that those
of the top panel.
 
\begin{figure}
   \centering
   \includegraphics[width=8.5cm]{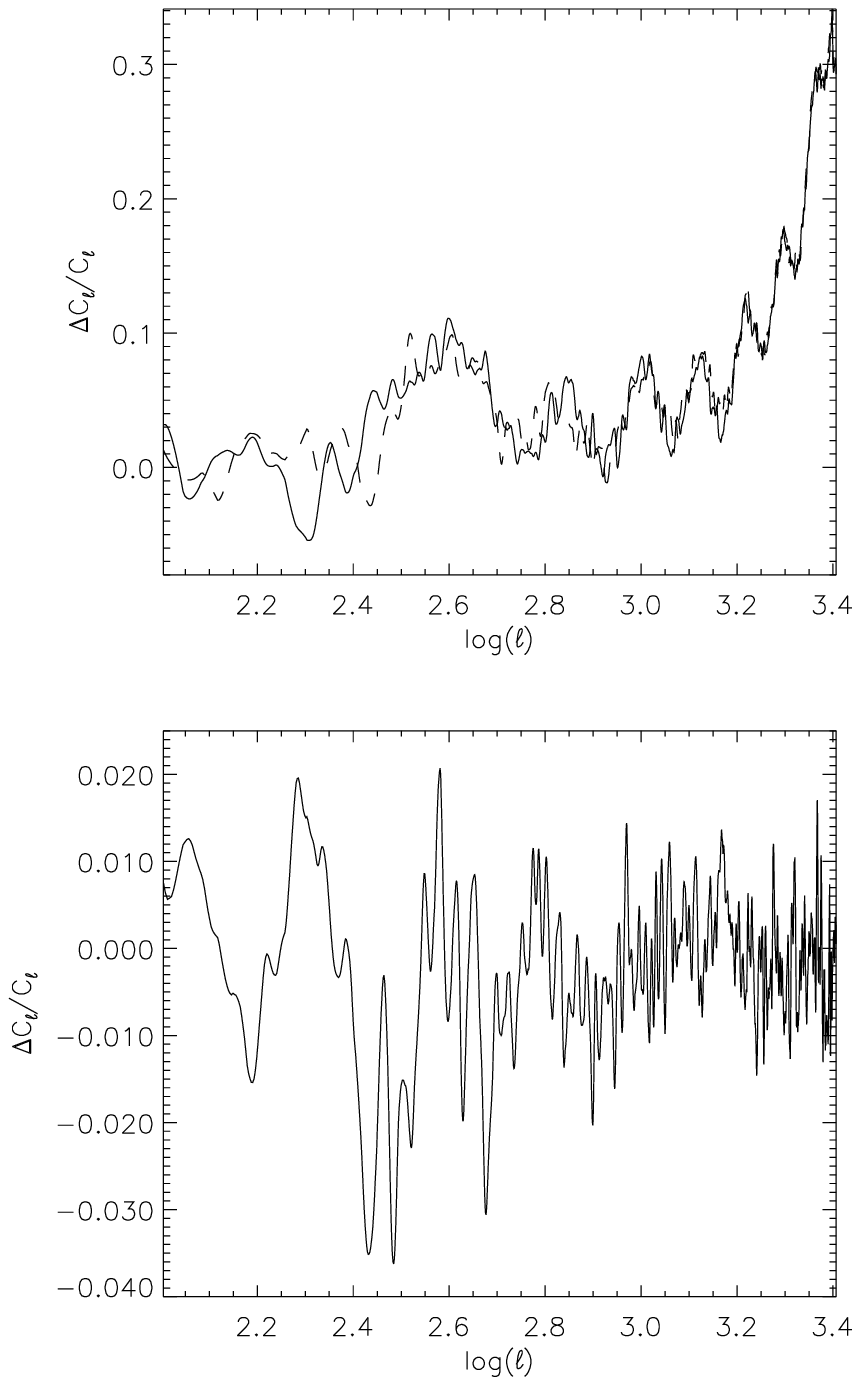}
      \caption{Solid and dashed lines of the top panel give
      the relative deviations between the theoretical spectrum and
      those corresponding to the solid and
      dashed lines of the middle panels of Fig.~\ref{Fig1}. 
      Bottom panel gives
      the residual deviations defined in the text. 
      }
         \label{Fig2}
   \end{figure}

In order to give quantitative estimates, some 
relative deviations $\Delta C_{\ell}/C_{\ell}$ are
calculated for appropriate pairs of spectra.
These deviations are presented in Fig.~\ref{Fig2}.
The top panel of this Figure shows the
deviations of the spectra $C_{\ell}^{1}$  and
$C_{\ell}^{2}$ of Fig.~\ref{Fig1} with respect to the true $C_{\ell}$
coefficients. In both cases,
quantities $\Delta C_{\ell}/C_{\ell}$ are similar,
which suggests a systematic effect (deviation with respect
to the true spectrum). Furthermore, the maxima and minima
of $\Delta C_{\ell}/C_{\ell}$ seems to be associate
to the minima and maxima in $\ell (\ell + 1) C_{\ell}$,
respectively.
Finally, the bottom panel of
Fig.~\ref{Fig2} shows the relative deviations between 
$\tilde{C}_{\ell}$ and $C_{\ell}$, which are hereafter
called residual deviations.
The largest
residual deviations are now of a few percent for any $\ell$, and
the peaks of this curve do not correspond to maxima and minima
in the angular power spectrum. The systematic deviations have been 
ruled out, and the residual deviations depend on the  pair of
simulations under consideration.
%

In order to separate the pixelisation effect and 
the REPC, 
the above study has been repeated in two cases: (i)
the previous pixelisation is maintained, whereas
a 500-simulation is used; 
hence, the new coverage is 
ten times larger than the previous one.
For the sake of briefness, no Figures are presented.
We only describe our main conclusions: 
(1) The systematic deviations has kept almost unaltered and,
(2) after subtracting these deviations, the 
residual ones range in the interval (-0.07,0.07);
namely, they are a factor $10^{-1/2}$
smaller than those of the bottom panel of Fig.~\ref{Fig2}.
These facts suggest that the systematic effect
is due to pixelisation, not to partial coverage, whereas
after subtraction, the resulting effect is due to partial coverage
(it is the REPC proportional to  $f_{\mbox{sky}}^{-1/2}$). In the 
case (ii) a 50-simulation
is used (the initial coverage), and a new pixelisation with
128 nodes per edge is assumed. 
In such a case, the spectrum can be only obtained 
for $\ell < \ell_{\mbox{max}} = \pi / \Delta=1275$ and,
for these $\ell$ values we observe that:
(a) a systematic effect appears again, (b)
the maxima and minima
of $\Delta C_{\ell}/C_{\ell}$ are again associated

%
   \begin{figure}[!h]
   \centering
   \includegraphics[width=8.5cm]{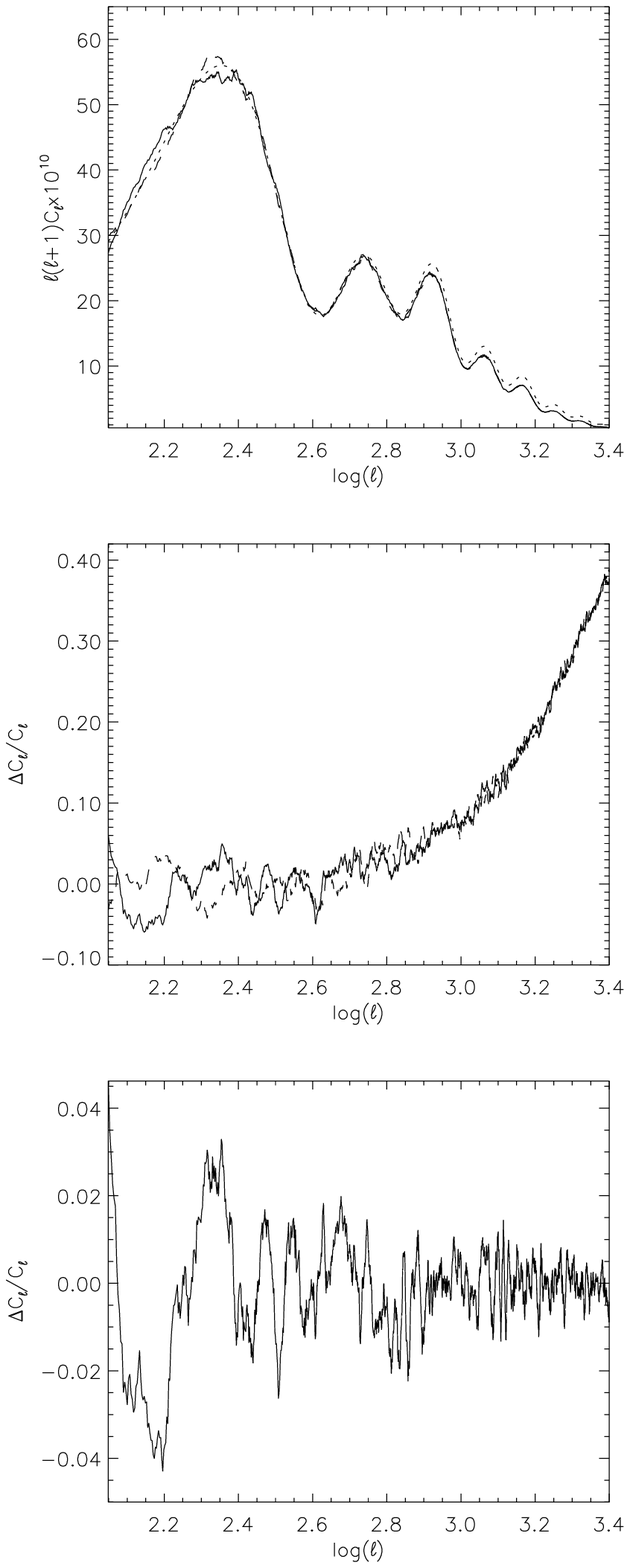}
      \caption{Top panel is the same as the middle panels 
               of Fig.~\ref{Fig1}, and middle (bottom) panel
               is the same as the top (bottom) panel of 
               Fig.~\ref{Fig2}. All the simulations are
               performed using HEALPIX with $\Delta = 3.435$}
         \label{nFig2}
   \end{figure}

to the minima and maxima in $\ell (\ell + 1) C_{\ell}$,
respectively, and (c) after subtraction of the systematic deviations, 
the residual deviations are very similar to those of the bottom 
panel of Fig.~\ref{Fig2}, as it should occur if such deviations are due to 
the unaltered partial coverage.

From the above considerations it follows that 
--after estimating and eliminating systematic 
effects due to 
pixelisation-- 
the spectrum extracted from fifty 
$18^{\circ} \times 18^{\circ}$ maps is 
very accurate for $100 < \ell < \ell_{\mbox{max}} $. 

The above study about the REPC and the pixelisation effect 
has been repeated using the HEALPIX package (see
Sect. 3). Two polar regions have been considered 
in each simulation ($f_{\mbox{sky}} \simeq
0.39$). The pixel size is $\Delta \simeq  3.435$; 
hence, we work 
with the same total sky coverage as in the case of squared 
patches, 
but with a different $\Delta$ value compatible with HEALPIX.
Results from two independent
simulations are presented in Fig.~\ref{nFig2}.
In the top panel, the solid and dashed lines show
the spectra obtained from these two simulations. These lines are
almost indistinguishable and,
consequently, they deviate 
almost the same with respect to the
dotted line (true spectrum). The
relative deviations (residual deviations) 
defined above are given in the middle (bottom)
panel. Residual deviations are similar
to those obtained using patches,
which is not surprising because the same sky coverage ($f_{\mbox{sky}} \simeq
0.39$)
has been assumed in both cases. 
Using simulations with 20 polar regions (sky coverage enhanced by
a factor ten), we have verified that the residual deviations 
reduce by a factor
$\sim 10^{1/2}$. We have also used two polar regions 
with ($f_{\mbox{sky}} \simeq 0.39$) and a 
greater pixel size $\Delta=6.87$ to get comparable 
residual deviations.
In short, using HEALPIX and polar regions (where the pixel
shapes are more irregular), the same qualitative results 
as in the case of squared patches have been obtained:
(1) the pixelisation effect 
appears to be very systematic, (2) 
the residual deviations correspond to the REPC and, (3)
the pixelisation effect and the REPC (residual deviations)
can be easily disentangled.
In next Sections, only methods 
based on squared patches are used.

%
   \begin{figure}
   \centering
   \includegraphics[width=8.5cm]{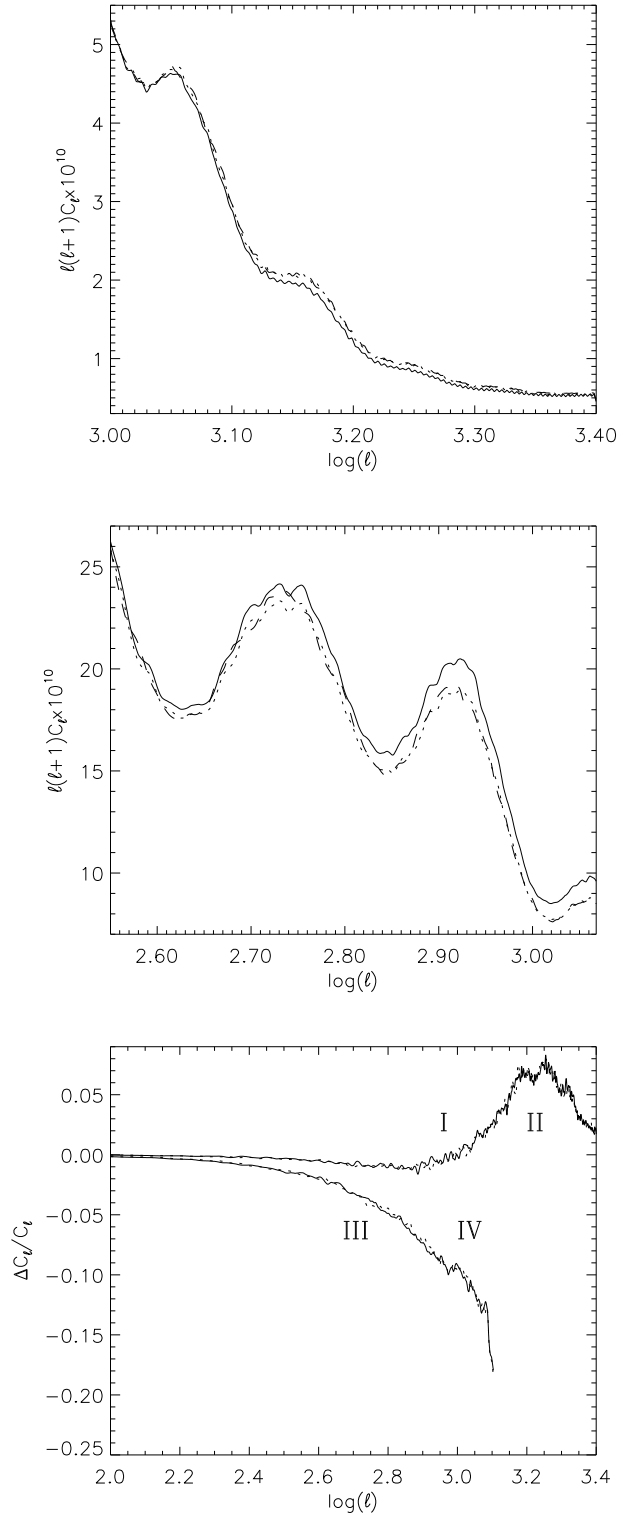}
      \caption{Top (middle) panel shows three spectra obtained from 
      50-simulations with $256 \times 256$ ($128 \times 128$) nodes
      per edge. In both panels, solid line is the spectrum obtained from
      a 50-simulation after
      smoothing with the circular associated beam and the other lines
      correspond to the same and to another independent 
      50-simulation after smoothing
      with the asymmetric non-rotating beam.
              Lines I and II (III and IV) of the bottom panel
              give the relative deviations between the solid line
              and each of the other two lines in the top (middle)
              panel.}
         \label{Fig3}
   \end{figure}

\subsection{Asymmetric non-rotating beam} 

The effect due to beam asymmetry is first
estimated in the absence of rotation. 
Maps are smoothed with both
the elliptical beam (which is 
oriented in the same way everywhere) and its 
circular associated one (see Sect. 2).
These beams lead to different averages at 
each node and, consequently, they produce
different alterations of the true angular power spectrum.
Results are presented in Fig.~\ref{Fig3}, 
whose top (middle) panel corresponds to maps
with 256 (128) nodes per edge. In these panels, 
the spectra are only shown in the $ \ell $ 
interval where the differences among them are more
relevant: $\ell > 1000$ ($\ell > 360$) for
256 nodes (128 nodes).
The solid line gives the 
spectrum after smoothing with
the spherical associated beam (which is denoted $C_{\ell}$ in this
Section). The
dotted and dashed lines give the spectra
obtained from two independent 50-simulations 
after smoothing with the 
nonrotating asymmetric beam (these spectra are denoted 
$C^{1}_{\ell}$
and $C^{2}_{\ell}$ along this Section).
These two lines are very similar and, consequently, 
they deviate almost the same with respect to the solid line.
This means that the effect due to asymmetry (without rotation)
is a very systematic one. The relative deviations between 
$C_{\ell}$ and $C^{1}_{\ell}$ and between 
$C_{\ell}$ and $C^{2}_{\ell}$
measure the effect 
of beam asymmetry without rotation. These 
deviations are displayed in the bottom panel
of Fig.~\ref{Fig3}, where curves I and II (III and IV) 
correspond to the $C^{1}_{\ell}$ and $C^{2}_{\ell}$
spectra displayed in the top (middle) panel. 
For the $256 \times 256$ pixelisation (top panel), the deviations 
are smaller than $\sim 8$ \% all along the 
interval (100, $\ell_{\mbox{max}}$), whereas 
for the $128 \times 128$ pixelisation (middle panel), 
these deviations
are greater than $\sim 8$ \% for $\ell > 700$, reaching
values close to $\sim 20$ \% near $\ell = \ell_{\mbox{max}}$.
There is a dependence on 
the pixelisation in the sense that, the smaller 
the pixel size the smaller the relative 
deviations (in the part of the spectrum common 
to both pixelisations). 
Curves I and II  
are very similar, and the same
occurs with curves III and IV, which 
means that the effect under consideration is actually systematic. 
If the systematic
effect is subtracted as in Sect. 4.1; namely,
if the quantities $D_{\ell}$, the corrected 
spectrum $\tilde{C}_{\ell}$, and the residual 
deviations are calculated, these last deviations 
are very similar 
to those showed in the bottom panel of Fig.~\ref{Fig2},
(associated to the REPC
in Sect. 4.1); they are oscillations having amplitudes
of a few per cent. 
Indeed, when 500-simulations are considered,
the amplitude of the residual deviations appear divided by
$10^{1/2}$ as it is expected in the case of any
deviation due to partial coverage (see Eq. (\ref{knox})).  
Finally, the deviations between ${C}_{\ell}$
and ${C}^{1}_{\ell}$ (or ${C}^{2}_{\ell}$) decrease as the 
assumed level of beam asymmetry does. 

%
   \begin{figure}[!h]
   \centering
   \includegraphics[width=8.5cm]{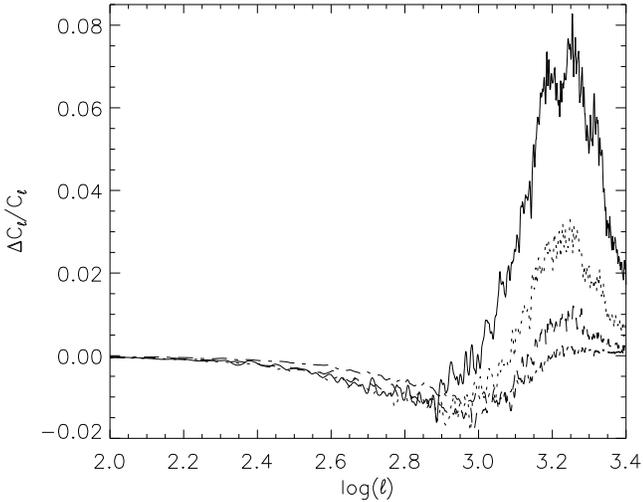}
      \caption{Relative deviations between the spectrum
      obtained from a
      50-simulation --with $256 \times 256$ nodes per edge-
      after smoothing with 
      the circular associated beam, and that obtained from the 
      same 50-simulation after smoothing with the asymmetric
      non-rotating beam. From top to bottom, 
      curves correspond to 
      different elliptical beams, in which the ratio
      $\sigma_{\phi} : \sigma_{\theta}$ is 10:6, 10:7, 10:8,
      and 10:9}
         \label{Fig4}
   \end{figure}

In Fig.~\ref{Fig4},
these deviations are shown for the 
$256 \times 256$ pixelisation and
different
elliptical beams. All these beams have 
$\sigma_{\phi}=10^{\prime}$, whereas 
$\sigma_{\theta}$ takes on the values 
$6^{\prime}$, $7^{\prime}$, $8^{\prime}$, and $9^{\prime}$. 
The decreasing has been also tested for
the $128 \times 128$ pixelisation but no Figure is 
presented by the sake of briefness.

\subsection{Asymmetric rotating beam} 

Various types of rotation strategies --of the 
asymmetric beam-- are now
introduced with the essential aim of estimating 
the rotation effects on the resulting angular power
spectra. Of course, these effects are deviations 
with respect to the spectrum
obtained --for the same asymmetric beam-- 
in the absence of rotation (${C}_{\ell}$ quantities
in this Section).
Two kinds
of rotation strategies are considered: 
in the first one, the beam orientation in
each pixel is that corresponding to a beam that does not
rotate around its centre, but it describes a big circle on 
the sky with and aperture angle of $85^{\circ}$. This 
first case mimics PLANCK observational strategy and
it is hereafter named Systematic Rotation (SR);
in the second strategy, the angle defining the 
beam orientation in 
each pixel is assumed to be a
random uniformly distributed variable; hereafter,
Random Rotation (RR).  
In each case, the rotating beam smoothes
fifty $18^{\circ} \times 18^{\circ}$ regions of
the sky which are distributed without overlapping 
and with random orientations. In this Section, 
$C^{1}_{\ell}$ and $C^{2}_{\ell}$ stand for two
spectra obtained from independent 50-simulations after 
a smoothing based on one of the above rotation strategies
(in each case, this strategy must be explicitly chosen).
   \begin{figure*}
   \centering
   \includegraphics{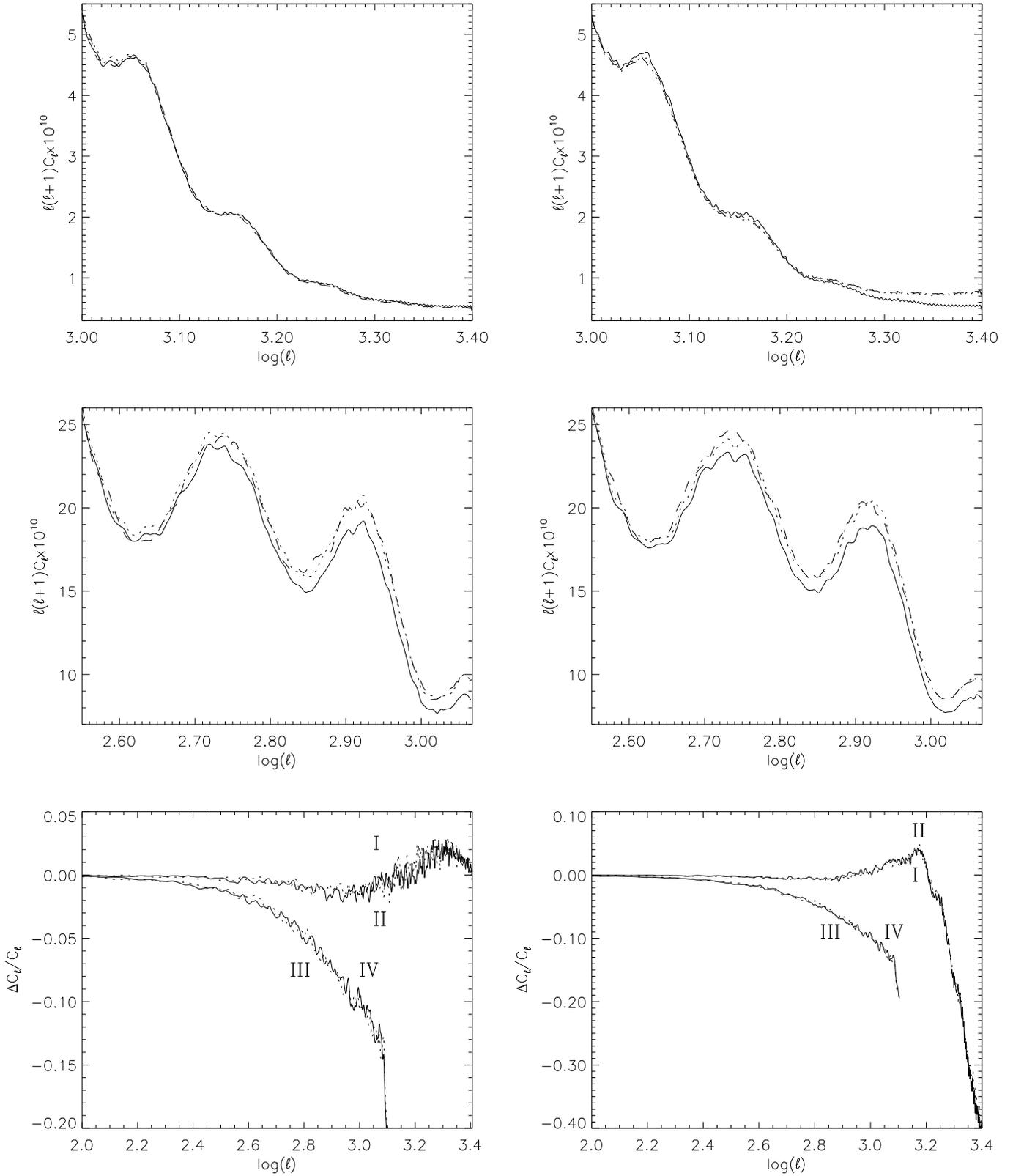}
   \caption{Top left panel shows three spectra
   for $\ell > 1000$: Solid line is the spectrum obtained from 
   a 50-simulation with $256 \times 256$ nodes per edge 
   after smoothing with the asymmetric 
   non-rotating beam. 
   Dashed and dotted lines are the spectra obtained
   from the same simulation and another independent one, 
   after smoothing with 
   the same asymmetric beam following the SR
   strategy defined in text. Top right panel is the same 
   as the top left one for the RR strategy. 
   Middle panels are as the
   top ones but simulations have $128 \times 128$ nodes per edge.
              Lines I and II (III and IV) of the bottom panel
              gives the relative deviations between the solid line
              and each of the other two lines in the top (middle)
              panel.}
              \label{Fig5}%
    \end{figure*}
Each of the fifty $18^{\circ} \times 18^{\circ}$ regions is 
smoothed only one time
with the asymmetric rotating beam.
That is a good procedure 
for experiments 
where the beam orientation 
varies from pixel to pixel (according to SR and RR strategies), 
but measurements are always performed with the same
orientation in each pixel. 
Results are displayed in Fig.~\ref{Fig5}, where
left (right) panel corresponds to SR (RR)
strategy.
Left and right panels have the same structure, which
is identical to that of Fig.~\ref{Fig3}; hence, 
on account of the definitions of 
$C_{\ell}$, $C^{1}_{\ell}$ and $C^{2}_{\ell}$ given in
this Section, 
the meaning of each line (and mark type) is known
and we can begin with the necessary discussion.
Top panels ($\Delta = 4.7^{\prime}$) show very different 
behaviours for $\ell > 1600$; while the RR produces an
important deviation of the dotted and dashed lines (two 
independent simulations)
with respect to the solid one (no rotation)
for $\ell > 1600$, the SR does not produces 
such a deviation. 
The deviations are described by
quantities $\Delta C_{\ell} / C_{\ell}$, which are
displayed in curves I and II (bottom panel).
Although these curves correspond to two independent 
50-simulations, they 
are very similar, which means that the effect 
under consideration is very systematic; in fact,
the residual deviations (after subtraction of the 
systematic effect) are of a few percent as in Sect. 4.1
(see the bottom panel of Fig.~\ref{Fig2}).

Middle panels show that, for $\Delta = 9.4^{\prime}$),  
SR and RR produce similar effects in all the interval
(100, $\ell_{\mbox{max}}$), where $\ell_{\mbox{max}} = 1275$.
For these $\ell$ values,
top panels also show similar effects for SR and RR. 
Relative errors 
are given in curves III and IV of the bottom panels.
The effect is again a systematic one and the
residual deviations has the same characteristics 
as in the case $\Delta = 4.7^{\prime}$
(top panels), which is easily understood taken 
into account that top and middle panels correspond
to the same partial coverage.

Finally, a pair of numerical experiments whose results 
are not displayed in
Figures: (i) we have increased the coverage using
500-simulations and
the main effect is that the new spectra are very
similar to those of Fig.~\ref{Fig5}, but they have much smaller 
high frequency oscillations (all the curves are much
more smooth), which is associated to the 
decreasing of the residual deviations
(the same behaviour have been already 
found in Sects. 4.1 and 4.2), and (ii) 
in the second experiment, each of the fifty 
$18^{\circ} \times 18^{\circ}$ regions 
have been smoothed $N_{\mbox{c}}$ times. 
Now, the orientation varies from pixel to pixel in 
each smoothing (according to SR and RR strategies)
and, furthermore, the orientation 
changes inside a given pixel from smoothing to
smoothing. Thus, we account for the fact that, 
in the framework of some experiments
(PLANCK mission),
each pixel is observed $N_{\mbox{c}}$ times with different
orientations.
For the SR strategy (which is similar to that 
of PLANCK), the minimum $N_{\mbox{c}}$ value is taken to be 
equal to the entire part of the number $N_{\mbox{c}}$ estimated
in Sect. 1; hence, we take $N_{\mbox{c}}=3$ for $\Delta = 4.7^{\prime}$  
and $N_{\mbox{c}}=6$ for $\Delta = 9.4^{\prime}$; we also take 
the values $N_{\mbox{c}}=30$ (for $\Delta = 4.7^{\prime}$) and 
$N_{\mbox{c}}=60$ (for $\Delta = 9.4^{\prime}$) in order to consider 
the existence of multiple detectors in current or 
future experiments.
In the case of the 
RR strategy, we take the same $N_{\mbox{c}}$-values.
Even for the greatest $N_{\mbox{c}}$ values, 
no appreciable differences are observed with respect to the 
case $N_{\mbox{c}}=1$. The same residual deviations and high frequency
oscillations appear in all the cases; 
this result is not surprising taking 
into account that: (1) the partial coverage is that 
of a 50-simulation whatever the $N_{\mbox{c}}$ value may be and,
consequently, the residual deviations do not decrease 
as we smooth the same 50-simulation various times and,
(2) the effect produced by beam rotation 
is a systematic one which is obtained very accurately
from the first smoothing of the 50-simulation. 
This means that, if a second smoothing is performed, 
the resulting spectrum must be very similar to the first one, 
and the average spectrum should be also almost 
identical to that of each smoothing (with no 
cancellation of the contribution due to the REPC);
this means that, if the coverage is large 
(very systematic deviations), the $N_{\mbox{c}}$ value
is irrelevant.

\subsection{Including galactic foregrounds}  

In previous Sections, the effects produced by beam asymmetry 
and rotation have been described in detail for 50-simulations 
of the CMB signal. These effects appear to be 
rather small and very systematic. 
Is that due to
the homogeneous and isotropic
statistical character of the CMB temperature distribution?

%
   \begin{figure}[!hp]
   \centering
   \includegraphics[width=8.5cm]{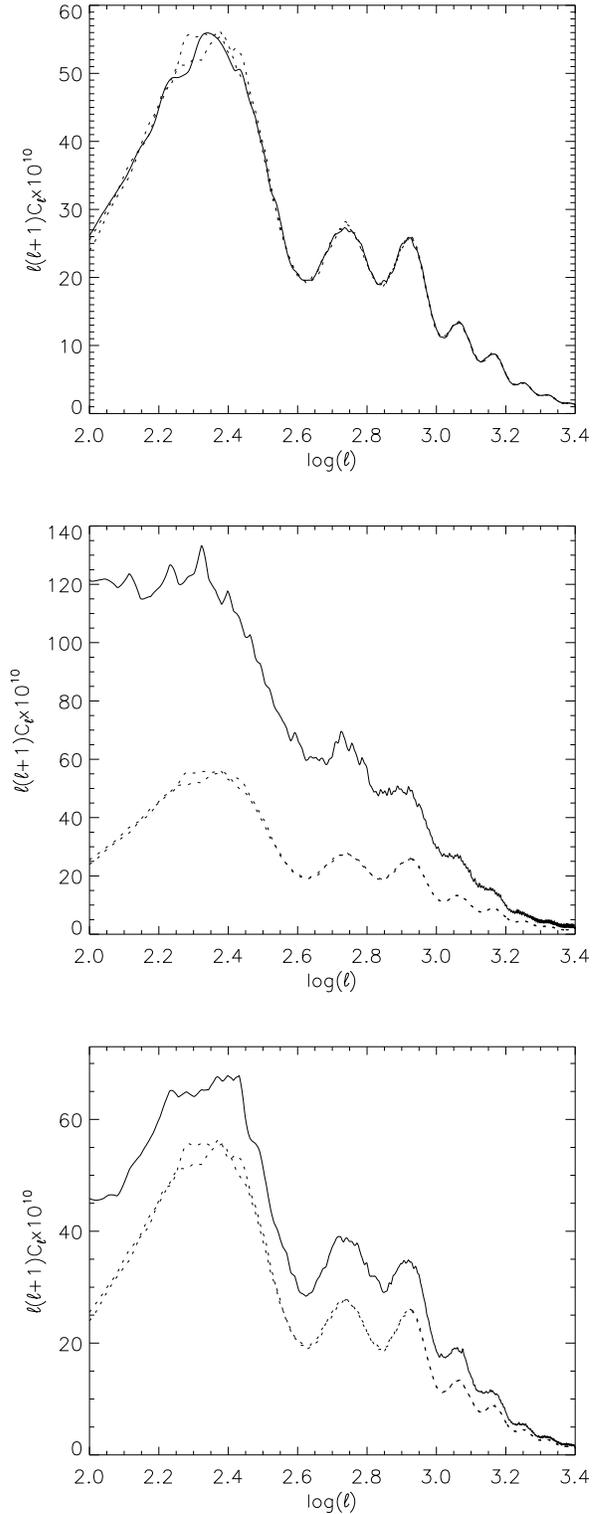}
      \caption{In all panels, solid line shows the
      spectrum of the CMB plus synchrotron and dust radiation 
      from a part of the Milky Way, and dotted and dashed lines 
      are the spectra extracted from two independent 50-realizations
      of pure CMB. No beam has been used at all. 
      All the spectra have been found from fifty 
      $18^{\circ} \times 18^{\circ}$ regions of the sky 
      with $256 \times 256$ nodes per edge. Top, middle
      and bottom panels corresponds to the regions of
      the sky G4--G5, G1--G2, and G3--G4, respectively.
      These regions are defined in the text.
              }
         \label{Fig6}
   \end{figure}
%

   \begin{figure*}[!hp]
   \centering
   \includegraphics{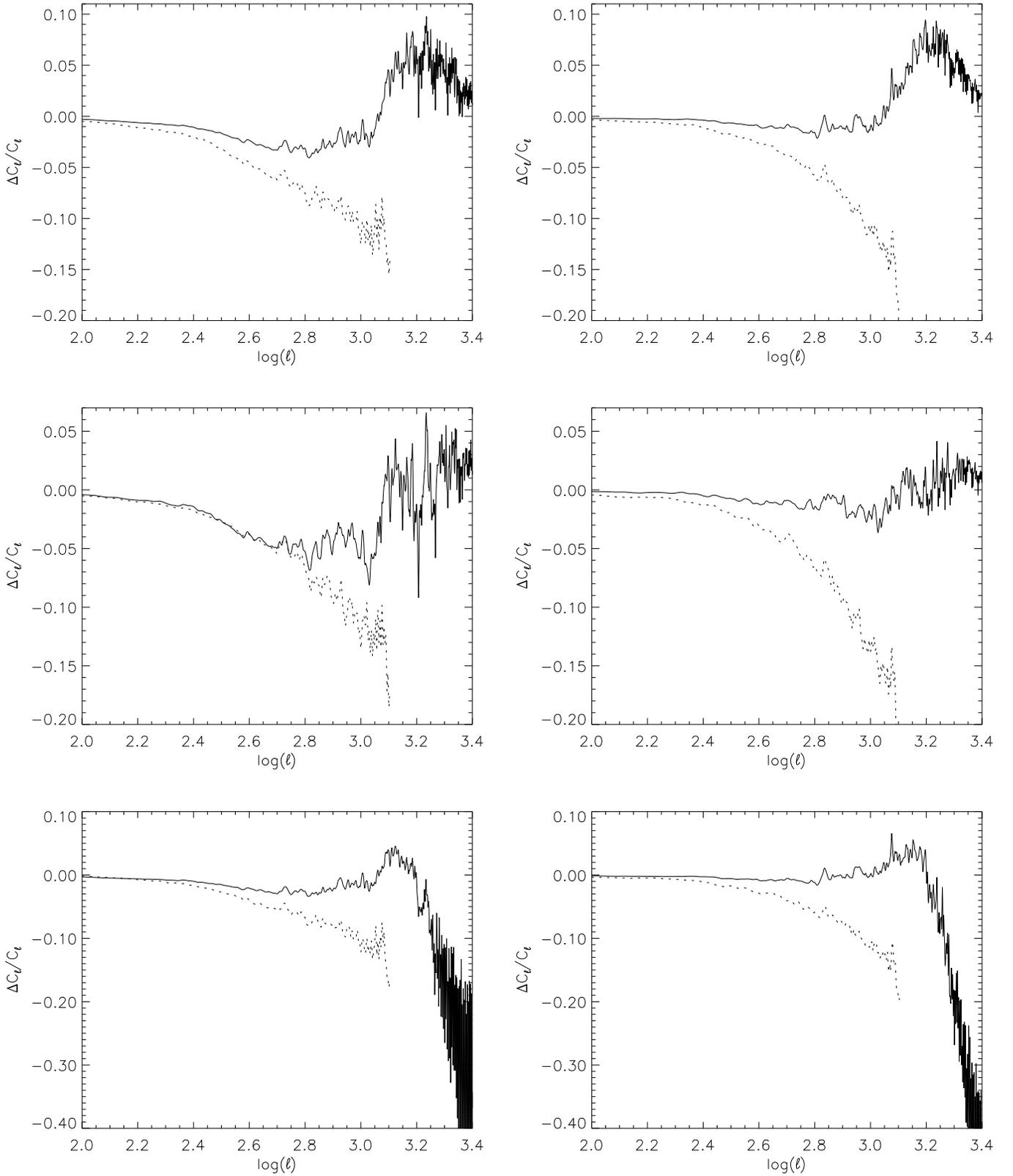}
   \caption{Left (right) panels correspond
   to 50f-simulations --see text--
   containing synchrotron and dust radiations 
   from regions G1 and G2  (G3 and G4) defined in the text.
   Top panels show deviations between the spectra extracted 
   from a 50f-simulation in two cases: after smoothing with 
   the circular associated beam, and after using the 
   non-rotating asymmetric beam. Middle (bottom) panels
   also show deviations, but the compared spectra are
   obtained after smoothing with the non-rotating asymmetric beam 
   and with the rotating one in the case of SR (RR).
   Solid (dotted) lines correspond to $256 \times 256$
   ($128 \times 128$) nodes per edge.
   }
              \label{Fig7}%
    \end{figure*}

What happens with other types of distributions?
The CMB is contaminated by galactic and extragalactic 
foregrounds which are not homogeneous 
and isotropic 
statistical fields and, consequently, the following 
question arises:  
what can we say about 
asymmetry 
and rotation effects in the presence
of the most important galactic foregrounds?
Are these effects very different from those estimated in previous 
Sections?

The galaxy produces an unique 
temperature distribution on the sky at a given frequency,
which must be known (from observations) before possible subtraction.
We are interested in a frequency channel involving small galactic 
contamination. The chosen frequency is  
100 GHz. 
We have used synchrotron and dust maps of the full sky,
which were taken from ESA (the maps and their technical
description are freely available at
astro.estec.esa.nl). These maps were designed by
G. Giardino and P. Fosalba and they have HEALPIX structure 
and a pixel size $\Delta = 1.7^{\prime}$.
Using these maps and an
appropriate smoothing, we have built up one hundred and fifty
$18^{\circ} \times 18^{\circ}$ maps with
$\Delta = 4.7^{\prime}$ and the same number of
maps with $\Delta = 9.4^{\prime}$.  

50-Simulations of the CMB signal 
are obtained as in previous Sections, and groups of fifty
$18^{\circ} \times 18^{\circ}$ maps
with the galactic foregrounds are
appropriately selected (see below); then,
each of the $18^{\circ} \times 18^{\circ}$ CMB maps is 
superimposed to one of the 
$18^{\circ} \times 18^{\circ}$ foreground maps
to get a 50f-simulation (which includes the CMB signal 
plus synchrotron and dust from the Milky Way).
Then, the study of previous sections is performed 
on the resulting 50f-simulation.

In order to get $18^{\circ} \times 18^{\circ}$ maps
with the foregrounds, 
a cube is inscribed in the celestial sphere in such a
way that the centres of the faces are distributed as
follows: (i) centres 1 and 2 point towards the centre 
of the Milky Way and its opposite direction,
centres 3 and 4 point toward two opposite 
directions which are contained in
the galactic plane and are orthogonal to the direction 
joining centres 1 and 2 and, (iii) centres 5 and 6
point towards the two galactic poles.
Each of the six faces is divided in 25 regions, in
such a way that after projection on the sphere,
our 150 maps are almost squared and
have similar areas and a small overlapping.
The 25 maps corresponding to the face with centre at $i$
are hereafter named the 25-group $G_{i}$.

The first 50f-simulation is built up using the 
25-groups $G_{5}$
and $G_{6}$ localized around the galactic poles. 
The solid line in the top panel of Fig.~\ref{Fig6}
gives the spectrum corresponding to 
this 50f-simulation, which can be compared with the two
spectra showed in the pointed and dashed lines, which 
correspond to two independent 50-simulations without
galactic foregrounds. No beam has been considered at
all. The small differences among these curves appear as
a result of the REPC, and 
the galactic contribution contained in the continuous line is 
hidden by this dominant effect.

The second (third) 50f-simulation includes the 25-groups
$G1$ and $G2$ ($G3$ and $G4$); hence, 
the second (third) 50f-simulation includes 
the Milky Way centre (a part of the galactic disk). 
The effect 
of the galactic foregrounds is not negligible.
That can be seen in the middle (second 50f-simulation) 
and bottom (third 50f-simulation) panels of Fig.~\ref{Fig6},
where the spectra containing the galactic foregrounds 
(solid line) are very different from the spectra 
corresponding to 50-simulations of pure CMB 
(dotted and dashed lines). No beam has been used.

The same asymmetric beam and
rotational strategies as in previous sections have been 
used to smooth each 50f-simulation. 
Results are presented in Fig.~\ref{Fig7}.
Left (right) panels corresponds to the second (third)
50f-simulation. In all the panels, the solid (dotted) line
shows results from $256 \times 256$ 
($128 \times 128$) simulations.
All the panels give relative deviations between 
two spectra, which are obtained after
smoothing with: (i) the non rotating beam and
the circular associated one in the top panels,
(ii) the non rotating beam and the rotating one
with SR strategy in the middle panel and, 
(iii) the non rotating beam and the rotating one
with RR strategy in the bottom panels. 
The top panel measures the effect of beam asymmetry without 
rotation, this panel should be compared with the bottom 
panel of Fig.~\ref{Fig3}, which shows the same effect in the
absence of galactic foregrounds.
The middle (bottom) panel gives the effect due to 
SR (RR) strategy and, consequently, this panel must
be compared with the left (right) bottom panel of
Fig.~\ref{Fig5}. The comparisons show some appreciable 
differences produced by the presence of 
significant galactic foregrounds (see Fig.~\ref{Fig6}).
In Fig.~\ref{Fig7}, the high frequency oscillations 
are most important that in the Figures used
for comparison; particularly, for large $\ell$ 
values. That could be due to the presence of 
oscillations
in the angular power spectrum of the galactic foregrounds
(see Fig.~\ref{Fig6} for the same $\ell$ values).
For small and intermediate $\ell$ values, 
the foreground contributions
to the angular power spectrum are significant
(see Fig.~\ref{Fig6}) and, consequently, 
there are also 
appreciable differences between 
the $\Delta C_{\ell} / C_{\ell}$ deviations displayed in
Fig.~\ref{Fig7} and those of the Figures used for comparison 
(without foregrounds). In spite of the
fact that the assumed foregrounds are not 
homogeneous and isotropic statistical fields, 
the mentioned differences are
not very large. They  
are more important in the case of
the left panels, in which, directions pointing
close to the Milky Way centre have
been considered. 
For the first 50f-simulation (galactic poles) the
$\Delta C_{\ell} / C_{\ell}$ deviations are not presented in Figures
because no appreciable differences appear in this case with
respect to 50-simulations without foregrounds.

\section{Discussion}

In the absence of beam, the pixelisation effect and the REPC 
have been disentangled to conclude that
pixelisation produces very systematic deviations
with respect to the true angular power spectrum. This 
conclusion has been obtained using two very different 
methods for simulations and data analysis (see Sect. 3)

We have studied the deviations in the angular power spectrum produced 
by the rotation of an asymmetric beam. Two
rotation strategies 
have been considered. One of them (SR) is similar to that 
of future experiments as PLANCK. The second strategy
(RR) is very different from SR, and it
has been introduced for comparisons. 
Maps with and without the dust and 
synchrotron  
radiations from the Milky Way (at $100 \ GHz$) have been considered. 
In Sect. 4, the rotation effects 
corresponding to different cases have been described and
compared, now let us 
present some general comments. 
 
If radiation from the galaxy is not considered, the
most important conclusion is that
rotation effects are very 
systematic for any rotation strategy and $f_{\mbox{sky}} = 0.39$. 
They are so systematic that we can subtract 
the deviations appeared in a 50-simulation, 
from the spectrum of another one, to recover very well the
spectrum corresponding to the nonrotating beam 
(except for small deviations
which seem to be essentially due to the REPC).
Furthermore, the resulting effects depend on the
rotation strategy, in particular, for large $\ell$
values and, consequently, they must be estimated 
--using simulations-- in each particular case.

Radiation from the galaxy --which can be seen as a non 
homogeneous an non isotropic statistical field-- contributes
significantly to the observable signal, except in the 
case of the polar galactic regions (G5--G6). In 
the G1--G2 and G3--G4 cases, the effect of beam rotation 
is significantly, but not dramatically, different from 
that obtained in the
absence of foregrounds.

After the deviations in 
the angular power spectrum due to beam asymmetry and
rotation have been estimated and characterized
(the main goal of this paper) and,
after proving that beam effects are very systematic,
some practical applications can be easily outlined.

Take the CMB power spectrum corresponding
to a certain theoretical model of structure formation in 
a given universe, 
take also a model for the foregrounds, 
a pixelisation,
the asymmetric beam for a given frequency, and the rotation 
strategy of an experiment with a
large enough coverage (i.e. PLANCK), and then, 
use a simulation --as the 50-simulations of this paper or similar--
to find 
the spectrum $C_{\ell}^{^{\mbox{CMB}}}$ after smoothing 
with the asymmetric rotating beam.
Repeat the simulation a large enough number of
times and verify that the resulting $C_{\ell}^{^{\mbox{CMB}}}$ spectra
are similar in all cases (systematic character).
Finally, use the deviations among the resulting spectra to
assign an error bar to $\langle C_{\ell}^{^{\mbox{CMB}}} \rangle$.
Use these data --obtained from simulations-- to 
answer the following
question: Is the theoretical model under
consideration compatible with the 
observational data from the experiment? In order to find
the answer, the observational data could be 
analyzed as follows: (i) Eliminate a part of the 
instrumental noise 
using an appropriate method (wavelets, 
Fourier transform, and so on), (ii) Separate components 
(CMB, synchrotron from our galaxy, and so on) taking
into account the frequency dependences, but 
keeping beam smoothings unaltered (usually, the beams
are eliminated at this stage under simplifying assumptions and
without considering rotation),
(iii) use the map of the CMB component 
--which has already been separated from foregrounds--
to extract the 
experimental spectra, $C_{\ell}^{^{\mbox{CMB}}}(exp)$, and finally
(iv) compare $C_{\ell}^{^{\mbox{CMB}}}$ with
$C_{\ell}^{^{\mbox{CMB}}}(exp)$ and study if these spectra can be 
identified taking into account the error bars. If they
can, the theoretical model is compatible with observations.
Note that --at the last step of the process-- 
we compare a simulated spectrum with an observational one,
and note also that both spectra are obtained from maps 
which have been 
smoothed with the same rotating asymmetric beam; hence,
the proposed method for data analysis includes beam rotation,
treating it (after verification) as the source of a very 
systematic effect.
Of course, this method has been only outlined, and much more
work would be necessary before implementation.


\begin{acknowledgements}
Part of this work was 
supported by the Spanish MCyT (project AYA2000-2045). 
Some calculations
were carried out on a SGI Origin 2000s at the Centro de Inform\'atica
de la Universidad de Valencia. We wish to thank an anonymous 
referee for constructive comments.
\end{acknowledgements}

\end{document}